\documentclass[12pt]{article}
\begin{document}
\newfont{\elevenmib}{cmmib10 scaled\magstep1}%
\newcommand{\tabtopsp}[1]{\vbox{\vbox to#1{}\vbox to12pt{}}}

\newcommand{\preprint}{
            \begin{flushleft}
   \elevenmib Yukawa\, Institute\, Kyoto\\
            \end{flushleft}\vspace{-1.3cm}
            \begin{flushright}\normalsize  \sf
            YITP-00-31\\
           {\tt hep-th/0005278} \\ May 2000
            \end{flushright}}
\newcommand{\Title}[1]{{\baselineskip=26pt \begin{center}
            \Large   \bf #1 \\ \ \\ \end{center}}}
\newcommand{\Author}{\begin{center}\large \bf
            S.\, P.\, Khastgir and R.\, Sasaki
     \end{center}}
\hspace*{0.7cm}%
\newcommand{\Address}{\begin{center} \it
            Yukawa Institute for Theoretical Physics, Kyoto
            University,\\ Kyoto 606-8502, Japan
      \end{center}}
\newcommand{\Accepted}[1]{\begin{center}{\large \sf #1}\\
            \vspace{1mm}{\small \sf Accepted for Publication}
            \end{center}}
\baselineskip=20pt

\preprint
\thispagestyle{empty}
\bigskip
\bigskip
\Title{Liouville Integrability of Classical Calogero-Moser Models}
\Author

\Address
\vspace{1.5cm}

\begin{abstract}
Liouville integrability of classical Calogero-Moser models is proved
for models based on any root systems, including the
non-crystallographic ones.
It applies to all types of elliptic potentials, {\em i.e.} untwisted
and twisted together with their degenerations (hyperbolic, trigonometric and
rational), except for the rational potential models confined by a harmonic
force.
\end{abstract}
\bigskip
\bigskip
\bigskip

In this note we demonstrate the Liouville integrability of  classical
Calogero-Moser models \cite{CalMo} based on any root systems
\cite{OP1} including  the non-crystallographic ones.
This applies to  models with all possible forms of the potentials,
in particular, various types (untwisted and twisted) of elliptic potentials,
except  for the one with rational potential in the confining harmonic force.
The quantum version of this result, restricted to non-elliptic potential
cases,
is reported in \cite{kps}.

 The proof of Liouville integrability is obtained by
combining two known facts:
The first is the universal Lax pair of Bordner-Corrigan-Sasaki \cite{bcs2}.
This is obtained by unifying the known Lax pairs of
various types \cite{bcs1},
\cite{DHoker_Phong}
in terms of representations of Coxeter group, which is the
symmetry group of Calogero-Moser models.
The universal Lax pair provides
a complete set of integrals of motion for each Calogero-Moser model
with any potential and based on any root system.
The second ingredient is a Theorem by
Olshanetsky and Perelomov \cite{OP3,OP1} on the structure of the
conserved quantities of Calogero-Moser models in general.
The latter simply asserts that for the conserved
quantities of Calogero-Moser
models \(\{Q_n\}\) satisfying certain conditions to be listed below,
any Poisson brackets among them \(\{Q_n,Q_m\}\) must vanish.

First let us recapitulate the basic ingredients of
the Calogero-Moser models, the Hamiltonian, the Lax pair and conserved
quantities in order to set the stage and to introduce notation.
 A Calogero-Moser model is a   Hamiltonian system associated with a root
system
\(\Delta\) of rank \(r\).
Quantum versions of these models are also integrable, at least
for degenerate potential functions \cite{kps} for any choice of \(\Delta\).
The dynamical variables are the coordinates
\(\{q_{j}\}\) and their canonically conjugate momenta \(\{p_{j}\}\), with
the Poisson brackets
\begin{equation}
  \{q_{j},p_{k}\}=\delta_{jk},\qquad \{q_{j},q_{k}\}=
  \{p_{j},p_{k}\}=0,\quad j,k=1,\ldots,r.
\end{equation}
These will be denoted by vectors in \(\mathbf{R}^{r}\)
\begin{equation}
  q=(q_{1},\ldots,q_{r}),\qquad p=(p_{1},\ldots,p_{r}).
\end{equation}
The Hamiltonian for the Calogero-Moser model is
\begin{equation}
   \label{CMHamiltonian}
   \mathcal{H}(p,q) = {1\over 2} p^{2} +
   {1\over2}\sum_{\alpha\in\Delta_+}
   {g_{|\alpha|}^{2} |\alpha|^{2}}
   \,V_{|\alpha|}(\alpha\cdot q),
\quad \Delta_+: \mbox{set of positive roots},
\end{equation}
in which the real coupling constants \(g_{|\alpha|}\) and potential
functions
\(V_{|\alpha|}\) are defined on orbits of the corresponding
finite reflection (Coxeter, Weyl) group, {\it i.e.} they are
identical for roots in the same orbit.
The generic potential is elliptic.
It is given by Weierstrass' \(\wp\) function:
\begin{equation}
   V_{|\alpha|}(\alpha\cdot q)=\wp(\alpha\cdot q|\{2\omega_1,2\omega_3\}),
   \quad \mbox{for all roots},
   \label{simppot}
\end{equation}
which is called {\em untwisted} model.
Here the two standard periods of the \(\wp\)
function are explicitly displayed.
In the {\em twisted}
model \cite{Ino,DHoker_Phong,bcs1}  the form of the potentials depends on
the
length of the roots. For long roots, it is the same as above. For short
roots
\begin{equation}
   V_{S}(\alpha\cdot q)=
   \left\{\begin{array}{cc}
\wp(\alpha\cdot q|\{\omega_1,2\omega_3\}),
   & \mbox{for}\ B_r, C_r, F_4,\\[10pt]
\wp(\alpha\cdot q|\{{2\omega_1\over3},2\omega_3\}),
   & \mbox{for}\ G_2.
\end{array}
 \right.
\end{equation}
That is, one of the two periods is reduced to be one half for
those root systems
 with \(|\alpha_S|^2/|\alpha_L|^2=1/2\)
and to one third for \(G_2\) in which \(|\alpha_S|^2/|\alpha_L|^2=1/3\).
The other potentials, the hyperbolic, the trigonometric and
the rational ones are
obtained as one (two) period tends to infinity:
\begin{equation}
   V_{|\alpha|}(\alpha\cdot q)={a^2\over{\sinh^2 a\alpha\cdot  q}},\quad
  {a^2\over{\sin^2 a\alpha\cdot  q}},\quad
  {1\over{(\alpha\cdot  q)^2}},\quad a:\ const.
\end{equation}
This then ensures that
the Hamiltonian is invariant under reflections of the phase space
variables about a hyperplane perpendicular to any root
\begin{equation}
  {\cal H}(s_{\alpha}(p),s_{\alpha}(q))={\cal H}(p,q), \quad
   \forall\alpha\in\Delta
  \label{HamCoxinv}
\end{equation}
with the action of \(s_{\alpha}\) on any vector
\(\gamma\in{\bf R}^r\) defined by
\begin{equation}
   \label{Root_reflection}
   s_{\alpha}(\gamma)=\gamma-(\alpha^{\vee}\!\!\cdot\gamma)\alpha,\quad
   \alpha^{\vee}\equiv2\alpha/|\alpha|^{2}.
\end{equation}

Next we describe the Lax pair and the conserved quantities.
Let us choose a set of \(\mathbf{R}^{r}\)
vectors
\({\cal R}=\{\mu^{(k)}\in\mathbf{R}^{r},\ k=1,\ldots, d\}\),
which form a \(d\)-dimensional representation of the Coxeter group.
That is, they are permuted
under the action of the Coxeter group.
Then the
Lax operators are \(d\times d\) dimensional matrices
\begin{eqnarray}
   \label{LaxOpDef}
   L(p,q) &=& p\cdot\hat{H}+X(q),\qquad X(q)
   =i\sum_{\rho\in\Delta_{+}}g_{|\rho|}
   \,\,(\rho\cdot\hat{H})\,x_{|\rho|}(\rho\cdot
   q,(\rho^{\vee}\!\!\cdot\hat{H})\xi)\,\hat{s}_{\rho},
   \\ \nonumber
   M(q) &=&
   {i\over2}\sum_{\rho\in\Delta_{+}}g_{|\rho|}|\rho|^2\,y_{|\rho|}
   (\rho\cdot q,(\rho^{\vee}\!\!\cdot\hat{H})\xi)\,\hat{s}_{\rho},
\quad \mbox{with}\quad  y_{|\rho|}(u,w)
\equiv {\partial x_{|\rho|}\over {\partial u}} (u,w).
\end{eqnarray}
consisting of operators \(\{\hat{H}_j\}\) and \(\{\hat{s}_{\rho}\}\).
Their matrix elements are defined by:
\begin{equation}
   (\hat{H}_{j})_{\mu\nu}=\mu_j\delta_{\mu\nu},\quad
    (\hat{s}_{\rho})_{\mu\nu}=\delta_{\mu,s_\rho(\nu)}=
      \delta_{\nu,s_\rho(\mu)},\quad \mu,\nu\in{\cal R}.
\end{equation}
The \(L\) and \(M\) operators are Coxeter covariant:
\begin{equation}
   L\left(s_{\rho}(p),s_{\rho}(q)\right)_{\mu\nu}=
   L\left(p,q\right)_{\mu^\prime\nu^\prime},\
   M\left(s_{\rho}(q)\right)_{\mu\nu}=
   M(q)_{\mu^\prime\nu^\prime},\quad \mu^\prime\equiv s_{\rho}(\mu),
   \  \nu^\prime\equiv s_{\rho}(\nu),
\label{lCoxinv}
\end{equation}
and they contain a spectral parameter \(\xi\) \cite{Krichever} which
enters through  functions \(x_{|\alpha|}(u,w)\).
For the untwisted model it is ($\wp(u)\equiv -d\zeta(u)/du$ and
$\zeta(u)\equiv d\log\sigma(u)/du$)
\begin{equation}
   \label{Genl_A2_Soln}
   x_{|\alpha|}(u,w) = {\sigma(w-u)
   \over{\sigma(w)
   \sigma(u)}}, \quad \mbox{for all roots},
\end{equation}
in which the dependence on the standard periods is suppressed.
For  long roots in the twisted model, it is the same as above and
for short roots
\begin{equation}
   x_{S}(u,w)=
   \left\{
    \begin{array}{lc}
       {\sigma(w/2-u|\{\omega_{1},2\omega_{3}\})
       \over{\sigma(w/2|\{\omega_{1},2\omega_{3}\})
       \sigma(u|\{\omega_{1},2\omega_{3}\})}} &
       \mbox{for}\ B_r,C_r, F_4\\[10pt]
       {\sigma(w/3-u|\{2\omega_{1}/3,2\omega_{3}\})
       \over{\sigma(w/3|\{2\omega_{1}/3,2\omega_{3}\})
       \sigma(u|\{2\omega_{1}/3,2\omega_{3}\})}} & \mbox{for} \ G_2.
    \end{array}
   \right.
\end{equation}
Near the origin \(q=0\), or \(u=0\), the \(u\)
dependence of the various \(x\)
functions is universal:
\begin{equation}
   x_{|\alpha|}(u,w)={1\over u}\left(1-u\,\zeta_c({w/ c})\right),\quad
   c=1,2,3,\quad  |u|\ll1,
\quad \zeta_c(u)\equiv \zeta(u|\{2\omega_1/c,2\omega_3\}),
\end{equation}
which is the same as that of the rational potential except for the spectral
parameter dependence:
\begin{equation}
   x_{|\alpha|}(u,w)={1\over u}\left(1-{u\over w}\right),\quad
   \mbox{for rational potential}.
   \label{ratlim}
\end{equation}

\bigskip

The canonical equations of motion can be cast into
the Lax form
\begin{equation}
   \label{LaxEquation}
   \dot{L}=[L,M]
\end{equation}
which implies that the traces of \(L\) are  conserved:
\begin{equation}
Q_n(p,q)\equiv\mbox{Tr}(L^n(p,q))=
\sum_{\mu\in{\cal R}}(L^n)_{\mu\mu},\quad
{d\over{dt}}Q_n= \{Q_n,{\cal H}\}=0, \quad n=1,2,\ldots,.
\end{equation}
The independent ones appear at such power \(n\) that it is 1+{\em exponent}
of
the root system. This is summarised in the following Table I:
\begin{center}
    \begin{tabular}{||c|l||c|l||}
       \hline
        \(\Delta\)& \(f_j=1+exponent\) &\(\Delta\)& \(f_j=1+exponent\)\\
       \hline
       \(A_r\) & \(2,3,4,\ldots,r+1\) & \(E_8\)
       & \(2,8,12,14,18,20,24,30\) \\
       \hline
       \(B_r\) & \(2,4,6,\ldots,2r\) & \(F_4\) & \(2,6,8,12\) \\
       \hline
       \(C_r\) & \(2,4,6,\ldots,2r\) & \(G_2\) & \(2,6\) \\
       \hline
      \(D_r\) & \(2,4,\ldots,2r-2;r\) & \(I_2(m)\) & \(2,m\) \\
      \hline
      \(E_6\) & \(2,5,6,8,9,12\) & \(H_3\) & \(2,6,10\) \\
      \hline
      \(E_7\) & \(2,6,8,10,12,14,18\) & \(H_4\) & \(2,12,20,30\) \\
      \hline
    \end{tabular}\\
 \bigskip
 Table I: The degrees \(f_j\) in which independent Coxeter
invariant polynomials exist.
\end{center}
Thanks to the availability of various representations for the Lax pair,
a complete set of independent conserved quantities \(\{Q_n(p,q)\}\) are
obtained  as traces of certain powers of \(L\). The independence of the
conserved quantities can be easily verified  by considering the free limit,
{\em i.e.}
\(g_{|\rho|}=0\).
They have the following properties:
\begin{enumerate}
\item
Coxeter invariance as a consequence of (\ref{lCoxinv}):
\begin{equation}
    Q_n(s_{\rho}(p),s_{\rho}(q))=Q_n(p,q),\quad \forall\rho\in\Delta.
\end{equation}
\item
\(Q_n(p,q)\) is a homogeneous polynomial of degree \(n\) in variables
\((p_1,\ldots,p_r,x_{|\alpha|}(\rho\cdot q,w))\), in which \(w\) is
proportional to the spectral parameter.
\item
Scaling property for those of rational potential models:
\begin{equation}
   {}^RQ_n(\kappa^{-1}p,\kappa q)=
   \kappa^{-n}\,{}^RQ_n(p,q)+\mbox{sub-leading
terms},
\end{equation}
as a consequence of the above point and (\ref{ratlim}).
\item
For the other types of potential, the asymptotic behaviour near the
origin:
\begin{equation}
   Q_n(p,q)={}^RQ_n(p,q)(1+{\cal O}(|q|)),\quad \mbox{for}\quad |q|\to0.
\end{equation}
\end{enumerate}
We need to show the vanishing of
\begin{equation}
J_{lm}\equiv [Q_l,Q_m],
\end{equation}
which is a polynomial in \(\{p\}\) of degree \(s\)
\begin{equation}
   s<l+m,
   \label{slesslm}
\end{equation}
since the leading powers in \(\{p\}\), \(p^l\) (\(p^m\))
in \(Q_l\) (\(Q_m\)),
having constant coefficients, commute with each other.
Let us decompose \(J_{lm}\) into the leading part and the rest:
\begin{equation}
   J_{lm}=J_{lm}^0+J_{lm}^{rest},\quad
   J_{lm}^0=\sum c^{j_1,\ldots,j_s}(q)p_{j_1}\ldots p_{j_s}
\end{equation}
and \(J_{lm}^{rest}\) is a polynomial in \(\{p\}\) of degree less than
\(s\).
The Theorem by Olshanetsky and Perelomov \cite{OP3} (Lemma 2, section 4)
states:
\begin{description}
\item{}
If the leading coefficients \(c^{j_1,\ldots,j_s}(q)\) are not constant, then
\(J_{lm}=0\).
\end{description}
Thus  the involution is proved.
The argument goes as follows. From Jacobi identity and conservation
\(\{{\cal H}, Q_{l(m)}\}=0\), we obtain
\begin{equation}
   \{{\cal H},J_{lm}\}=0.
\end{equation}
Considering the explicit form of the Hamiltonian (\ref{CMHamiltonian}), the
leading ({\em i.e.\/} of degree \(s+1\) in \(\{p\}\)) part of \(\{{\cal
H},J_{lm}\}\) comes only from the free part
\[
   \{p^2,J_{lm}^0\}
\]
and it vanishes if the following conditions are satisfied:
\begin{equation}
   \sum_{\sigma}{\partial\over{\partial q_t}}c^{k_1,\ldots,k_s}(q)
   =0,
\label{cqeq}
\end{equation}
where the sum is taken over all permutations of indices
\(\sigma(t,k_1,\ldots,k_s)=(j_1,\ldots,j_{s+1})\).
In \cite{Bere} it is proved (Lemma 2.5, p. 407)
that the system (\ref{cqeq}) has
only polynomial solutions.
 For rational potential  models the
scaling property of the leading terms tells that
\(c^{k_1,\ldots,k_s}(\kappa
q)=\kappa^{s-l-m}c^{k_1,\ldots,k_s}(q)\). Since \(s<l+m\)
(\ref{slesslm}), it follows that the only polynomial solution  satisfying
the condition is the null polynomial. Thus we obtain
\(c^{j_1,\ldots,j_s}(q)=0\) \(\Rightarrow J_{lm}^0=0\) and \(J_{lm}=0\).
The same results follow for the other types of potentials by considering the
asymptotic behaviour for \(|q|\to0\). Thus the involution of all the
conserved quantities \(\{Q_n\}\) is proved for the models with rational,
hyperbolic, trigonometric and various types of elliptic potentials
for all root systems.

It would be interesting to relate this type of argument to
other approaches to the involution properties, for example, those of the
classical \(r\)-matrices for elliptic models by
Sklyanin and others \cite{AvTal}-\cite{skly}.

We thank  A.\,J.\, Pocklington for useful discussion.
This work is partially
supported  by the Grant-in-aid from  the   Ministry of Education,
Science and
Culture, priority area (\#707)  ``Supersymmetry and unified theory of
elementary particles". S.\,P.\,K. is supported by the Japan
Society for the Promotion of Science.

\end{document}